\documentstyle[12pt,aaspp4]{article}
\begin{document}

\newcommand{\up}[1]{\ifmmode^{\rm #1}\else$^{\rm #1}$\fi}
\newcommand{\zdot}{\makebox[0pt][l]{.}}
\newcommand{\upd}{\up{d}}
\newcommand{\uph}{\up{h}}
\newcommand{\upm}{\up{m}}
\newcommand{\ups}{\up{s}}
\newcommand{\arcd}{\ifmmode^{\circ}\else$^{\circ}$\fi}
\newcommand{\arcm}{\ifmmode{'}\else$'$\fi}
\newcommand{\arcs}{\ifmmode{''}\else$''$\fi}

\title{The ARAUCARIA project. Discovery of Cepheid Variables
in NGC 300 from a Wide-Field Imaging Survey
\footnote{Based on  observations obtained with the 2.2~m ESO/MPI
telescope at the European Southern Observatory
}}

\author{Grzegorz Pietrzy{\'n}ski}
\affil{Universidad de Concepci{\'o}n, Departamento de Fisica, Grupo de 
Astronomia, Casilla 160--C, 
Concepci{\'o}n, Chile}
\affil{Warsaw University Observatory, Al. Ujazdowskie 4,00-478, Warsaw, Poland}
\authoremail{pietrzyn@hubble.cfm.udec.cl}
\author{Wolfgang Gieren}
\affil{Universidad de Concepci{\'o}n, Departamento de Fisica, Grupo de 
Astronomia, Casilla 160--C, 
Concepci{\'o}n, Chile}
\authoremail{wgieren@coma.cfm.udec.cl}
\author{Pascal Fouqu\'e}
\affil{Observatoire de Paris-Meudon DESPA, F-92195 Meudon CEDEX, France}
\affil{European Southern Observatory, Casilla 19001, Santiago 19, Chile}
\authoremail{pfouque@eso.org}
\author{Frederic Pont}
\affil{Universidad de Chile, Departamento de Astronomia, Casilla 36D, 
Santiago, Chile}
\authoremail{Frederic.Pont@obs.unige.ch}

\begin{abstract}
We have conducted a survey for Cepheid variables in the Sculptor Group spiral 
NGC 300. Based on observations obtained
with the Wide-Field Camera at the 2.2 m ESO/MPI telescope during 29 nights 
spread over a 5.3 month interval,
117 Cepheids and 12 Cepheid candidates were found which cover the period range 
from 115 to 5.4 days.
 We present a catalog which provides 
equatorial coordinates, period, time of maximum brightness, and
intensity mean B and V magnitudes for each variable, and we show phased B and V 
light curves for all the Cepheids found. We also present the individual B and V 
observations 
for each Cepheid in our catalog.
During our search we rediscovered all 18 previously known Cepheids, and 
confirmed the Cepheid nature of 3 Cepheid candidates from the previous 
photographic survey of Graham. Star V4 
in Graham's list, classified by him as an eclipsing binary, 
turns out to be another Cepheid. We find very good agreement between our 
photometry and that obtained by Freedman et al. from ground-based CCD data for 
common stars. Using the earlier
data together with our new data, we were able to significantly improve the
periods for 15 Cepheids in our sample. One of the Cepheids with earlier 
observations shows evidence for a change of its 
period over the last 26 years. The Cepheids delineate the spiral arms of NGC 
300, and a couple of them
were detected very close to the center of the galaxy. From the color-magnitude 
diagram of NGC 300 constructed
from our data, we expect that our Cepheid detection is near-complete for 
variables with periods larger than
about 10 days.

We present plots of the PL relations in the B and V bands obtained from our 
data, which
clearly demonstrate the presence of a Malmquist bias for periods below about 10 
days. A 
thorough discussion of the distance to NGC 300 will be presented in a 
forthcoming paper
which will include the analysis of photometry in longer-wavelength bands.
\end{abstract}


\section{Introduction}
One of the most fascinating and challenging chapters of modern astrophysics has 
been the 
effort to establish the extragalactic distance scale with an ever increasing 
accuracy. Key
to this effort has been the use of the Hubble Space Telescope over the last 
years which has allowed to detect
and measure the brightness of Cepheid variables in some 20 nearby spiral 
galaxies and determine
their distances from the Cepheid period-luminosity relation. In fact, the HST 
Key Project
team on the extragalactic distance scale has just recently presented their final 
results
for the Cepheid distances to their selected galaxies (Freedman et al. 2001), 
from which secondary methods of
distance measurement have yielded an improved determination of the Hubble 
constant. However, in the HST Key Project
procedure to determine the Hubble constant there are still a number of 
systematic uncertainties which
prevent, at the present time, a truly accurate determination of this constant 
which is so
important to cosmology. The most nagging problem continues to be the distance to 
the LMC, whose Cepheids
are used as a fiducial to determine the relative distances to the other 
galaxies. The distance
to the LMC is currently the largest single contributor to the systematic 
uncertainty on ${\rm H}_{0}$
(Mould et al. 2000). Another problem, affecting the calibrating Cepheid samples 
in different galaxies, is the 
dependence of Cepheid absolute magnitudes on metallicity. Currently there are 
very controversial
claims about the importance of this effect for the determination of Cepheid 
distances
to galaxies, and a truly accurate empirical determination of the effect is 
urgently needed to further
reduce the uncertainty on the Hubble constant.

In spite of the very important progress achieved over the last years, not only 
by HST but by many other teams 
of researchers using ground-based telescopes, it seems clear that the remaining 
uncertainty in
the distance scale can only be significantly reduced if we succeed to make true 
progress in determining
the distance to the LMC, and a number of other nearby galaxies. To achieve this, 
we must learn more about the  
systematic uncertainties in the principal methods currently used to find these 
distances. The most promising stellar methods,
apart from Cepheid variables, use red clump giants, the tip
of the red giant branch, eclipsing binaries, and planetary nebulae as distance 
indicators. Another, relatively new
tool to determine extragalactic distances from individual stars is the wind 
momentum-luminosity relationship (WLR)
for blue supergiants (Kudritzki et al. 1999; Bresolin et al. 2001). Only when 
the true capabilities
and systematics of these various methods of distance determination are fully 
understood, the techniques properly 
calibrated, and agreement
on the distances to a number of nearby galaxies with different environmental 
properties from all these
methods is achieved, can we trust that the local calibration of the distance 
scale has been
finally settled. A working group at the Universidad de Concepci{\'o}n (including 
several of the authors
of this paper) has recently been established to work on the improvement of 
stellar distance indicators. We
call our project the ARAUCARIA project (making reference to a famous tree at our 
southern location), and we
hope to make a significant contribution to the improved calibration of the local 
extragalactic distance scale,
over the next years.

Our target to start with is the Sculptor Group spiral galaxy NGC 300. At a 
distance of about 2.0 Mpc (Freedman et al. 2001),
this  galaxy is sufficiently close to resolve its stellar 
populations. Since NGC 300 shows clear signs of recent,
massive star formation, it seemed an excellent target to discover a significant 
number of Cepheid variables.
Indeed, a relatively small number of Cepheids in NGC 300 had already been 
previously discovered by Graham (1984) with
photographic techniques at the CTIO 4 m telescope, and a subsample of these 
variables was later observed
with a ground-based CCD camera by Freedman et al. (1992). However, taking 
advantage of new wide-field imaging
capabilities, we hoped to find many more Cepheids and determine their properties 
with a higher accuracy than
what had been achieved in former work, a hope which is borne out by the results 
presented in this paper. In a
parallel project, we  are conducting a survey for blue supergiant stars in the 
disc of NGC 300 with the ESO
VLT telescope; first results have recently been presented in Bresolin et al. 
(2001). The blue supergiants will be used to
establish the (stellar) metallicity gradient in the disc of NGC 300 to perform a 
new and accurate empirical
calibration of the effect of metallicity on Cepheid luminosities, one of the 
principal goals of our current
project. They will also be used to improve the WLR calibration for blue 
supergiants. In addition, we have also secured
narrow-band images of the galaxy to conduct a complete survey for its planetary 
nebulae population and improve
the calibration of the PNLF method of distance determination.

The present paper is organized as follows. In section 2, we describe the 
observations from which the current
survey for Cepheids was conducted. In section 3 we describe the reduction 
techniques and calibration of the
images. In section 4, we present the catalog of newly discovered Cepheid 
variables in NGC 300 (variable stars
of different types will be presented in a forthcoming paper). In section 5, we 
compare our new data
to the ones for those Cepheids for which previous photometry exists from the 
work of Graham, and of Freedman et al. (1992).
A discussion of the results and conclusions are presented in section 6.

\section{Observations}
All observations presented in this paper have been collected with 
the ESO/MPG 2.2 m telescope at the  La Silla observatory in  
Chile. The telescope 
was equipped with the ESO mosaic Wide Field Camera (WFI) consisting of 
eight 2048 x 4098 pixel arrays. The CCDs were separated by gaps 
of 23.8 and 14.3 arcsec in right ascension and declination
directions, respectively. The total field of view was about 
34 x 33 arcmin with a scale of 0.238 arcsec / pixel.

Our observations started on July 31, 1999 and lasted until January 8, 2001.
During this period, we secured 150 B, 138 V, 49 R and 145 I-band images centered 
on NGC 300, on 29 different nights
allocated to our programme. Most of these nights were photometric, and the 
seeing ranged from 0.7 to 1.5 arcsec
during the vast majority of the exposures taken.
In order to obtain photometry for all stars from the whole  observed field 
including those located in the gaps between CCDs, the observations were carried 
out in a "dithering" mode. During a given night  
5 consecutive 360 s exposures, slightly shifted in right ascension and/or 
declination
with respect to each other, were made in a given filter. This technique 
also allowed us to obtain a better accuracy of the photometric measurements by 
evaluating a
mean from all observations taken during a given night through a given filter,
and help to recognize bad measurements caused by weather, blending, or presence 
of bad pixels. On about half of the nights we obtained full
observing sequences through BVRI filters, in the other nights we could only 
observe BVR or BVI
sequences. On a few nights, only BV sequences could be obtained.
Effectively, 23-30 epochs, depending on the brightness  of a 
given star and filter, were obtained. The journal of B and V observations which
were used to conduct the search for Cepheids is presented in Table 1.  

\begin{deluxetable}{ccc}
\tablecaption{Journal of Observations}
\tablehead{
\colhead{UT date}  & \colhead{Filter}  & \colhead{HJD}\\
}
\startdata

 1999 July 31    &      B  &    2451390.780  \nl
 1999 August 1   &      B  &    2451391.936  \nl
 1999 August 1   &      V  &    2451391.846  \nl
 1999 August 2   &      B  &    2451392.766  \nl  
 1999 August 3   &      B  &    2451393.819  \nl
 1999 August 3   &      V  &    2451393.847  \nl
 1999 August 10  &      B  &    2451400.907  \nl
 1999 August 10  &      V  &    2451400.806  \nl
 1999 August 14  &      B  &    2451404.807  \nl
 1999 August 14  &      V  &    2451404.867  \nl
 1999 August 18  &      B  &    2451408.812  \nl
 1999 August 18  &      V  &    2451408.840  \nl
 1999 August 31  &      B  &    2451421.882  \nl
 1999 August 31     &   V  &    2451421.861  \nl
 1999 September  3  &    B &    2451424.811   \nl
 1999 September  3  &    V &    2451424.839  \nl
 1999 September 10  &    B &    2451431.797  \nl 
 1999 September 10  &    V &    2451431.824  \nl
 1999 September 12  &    B &    2451433.793  \nl
 1999 September 12  &    V &    2451433.822  \nl
 1999 September 15  &   B  &    2451436.799  \nl 
 1999 September 15  &   V  &    2451436.828  \nl
 1999 September 17  &   B  &    2451438.804  \nl 
 1999 September 17  &   V  &    2451438.832  \nl
 1999 October 30    &   B  &    2451481.576  \nl   
 1999 October 30    &   V  &    2451481.603  \nl
 1999 November 2    &   B  &    2451484.575  \nl  
 1999 November 2    &   V  &    2451484.632  \nl
 1999 November 4    &   V  &    2451486.577  \nl
 1999 November 6    &   B  &    2451488.535  \nl 
 1999 November 6    &   V  &    2451488.563  \nl
 1999 November 9    &   B   &   2451491.555  \nl
 1999 November 9    &   V  &    2451491.580  \nl
\enddata
\end{deluxetable}
\begin{deluxetable}{ccc}

\setcounter{table}{0}
\tablecaption{Journal of Observations - Concluded}
\tablehead{
\colhead{UT date}  & \colhead{Filter}  & \colhead{HJD}\\
}
\startdata
 1999 November 14   &   B  &    2451496.506  \nl
 1999 November 14   &   V  &    2451496.560  \nl
 1999 November 16   &   B  &    2451498.554  \nl
 1999 November 16   &   V  &    2451498.582  \nl
 1999 December 4    &   B  &    2451516.584  \nl
 1999 December 4    &   V  &    2451516.611  \nl
 1999 December 7    &   B  &    2451519.560  \nl
 1999 December 7    &   V  &    2451519.585  \nl
 1999 December 9    &   B  &    2451521.561  \nl
 1999 December 9    &   V  &    2451521.587  \nl
 1999 December 12   &   B  &    2451524.564  \nl
 1999 December 12   &   V  &    2451524.591  \nl
 1999 December 14   &   B  &    2451526.539  \nl
 1999 December 14   &   V  &    2451526.564  \nl
 1999 December 16   &   B  &    2451528.593  \nl
 1999 December 16   &   V  &    2451528.566  \nl
 1999 December 19   &   B  &    2451531.561  \nl
 1999 December 19   &   V  &    2451531.590  \nl
 1999 December 31   &   B  &    2451543.536  \nl
 1999 December 31   &   V  &    2451543.562  \nl
 2000 January 2     &   B  &    2451545.624  \nl
 2000 January 4     &   B  &    2451547.549  \nl
 2000 January 4     &   V  &    2451547.601  \nl
 2000 January 9     &   B  &    2451552.534  \nl
 2000 January 9     &   V  &    2451552.564  \nl
\enddata
\end{deluxetable}

\section{Reductions and Calibrations}
Preliminary reductions (i.e. debiasing and flatfielding) were carried out in the 
standard way with the  IRAF\footnote{IRAF is distributed by the
National Optical Astronomy Observatories, which are operated by the   
Association of Universities for Research in Astronomy, Inc., under cooperative
agreement with the NSF.} package. In order to avoid possible variations of the 
PSF across the image each of the 8 chips was divided into 2 slightly overlapping 
subframes. Then profile photometry for all stars detected on all subframes was
performed  with the $DAOPHOT$ and $ALLSTAR$ programs. The PSF model was derived 
iteratively. First, 10 relatively bright and isolated stars were selected and 
the first approximation of the PSF model was calculated. In the next step we 
subtracted all neighbor stars with the $ALLSTAR$ program and derived the PSF 
model again.
The PSF model obtained after three such loops was finally adopted. The 
photometric data from each 
two subframes were finally tied together using stars located in the overlapping 
regions.

In order to prepare an unique system of identification of stars in the observed 
field, the best (i.e. obtained under the best seeing conditions) set of five V 
images was
selected and stacked together. The resulting "template" included all stars 
located 
in the gaps between chips and covered a region of about 34 x 33 arcmin (see Fig. 
1).
The reference 
list with coordinates of objects from the template image was prepared with the 
$DAOPHOT$
program. After rejecting evidently spurious detections due to saturated stars or 
wild pixels, the total number of objects we detected was about 32000. The pixel 
coordinates of all 
stars from a given field were transformed to the template coordinate system, and 
a
unique number corresponding to the numbering system from the reference list
was assigned to each object.

A detailed description of the adopted photometric calibration procedure, 
completeness 
tests, a comparison with previous results together with a presentation 
and discussion of CMDs will be presented in a forthcoming paper. 
Briefly, to calibrate the photometric magnitudes of all stars on the wide field 
images an extensive sequence of secondary standard stars, distributed over 
almost the whole observed area and spanning a broad range in magnitudes 
and colors, was set up (Pietrzy{\'n}ski et al. 2001).
The WFI photometry was transformed to the 
standard system using about 100 stars from this list.
Derived residuals usually  did not exceed 0.04 mag and 
did not show any dependence on color or brightness. 
The accuracy of the zero point of our WFI photometry is about
0.03 mag in the standard V and B bands.
We found good agreement (the mean differences were ${\Delta
V=0.008\pm0.028}$~mag, ${\Delta(B-V)=-0.018\pm0.031}$~mag) between our
photometry and that obtained by Walker (1995), for 22 common stars in both
samples. 

Taking into account the periods of the potential Cepheids, all consecutive 
observations obtained in a given filter, during a given night
(amounting to typically 30 minutes) still define practically the same phase in
the light curve, even for the shortest-period Cepheids we detected (see section 
4). For
that reason we decided to improve the accuracy of our Cepheid measurements, for 
a given
night and in a given filter, by adopting the mean magnitude from the individual
dithered observations.

Equatorial coordinates were calculated with the algorithm developed and used 
by the OGLE team (Udalski et al. 1998). Shortly, a FITS file slightly larger   
than our template was extracted from the Digital Sky Survey (DSS) images and all  
stars having more than 200 counts above the sky level were detected on it.   
The pixel (x,y) coordinates of the detected stars were transformed to equatorial 
coordinates, and then to (x',y') pixel coordinates on the plane tangential to 
the
celestial sphere at the center of our field. The resulting two sets of pixel  
coordinates were tied together using third order polynomials. The internal
accuracy of our transformation was about 0.3 arcsec.

\section{The Cepheid Catalog}
As a first step, all stars were  subjected to a period search using the AoV 
algorithm (Schwarzenberg-Czerny 1989). 
We searched for periods between 0.2 and 90 days, the latter value corresponding
to about a half of the time span of our observations.  For the detected 
Cepheids, 
the accuracy of the derived periods is about 0.5-1 
$\times 10^{-2} \times {\rm P}$, or roughly 1 percent of the period lengths. 
However, for some of the 
Cepheids, the ones with previous observations from Freedman et al. (1992) and/or 
Graham (1984),
we were able to determine much more accurate periods by combining the different 
sets of photometric
measurements (see chapter 5 for more details).

In order to distinguish Cepheid variables from other types of
variable stars we used the following selection criteria: 

\begin{displaymath}
\begin{array}{l}
\mbox{1. Stars must exhibit the typical asymmetrical Cepheid-like shape of the 
light curves} \\ 
\mbox{2. ${\rm A}_{\rm B} > 0.4$ mag} \\
\mbox{3. ${\rm A}_{\rm B} > {\rm A}_{\rm V}$} \\
\mbox{4. 0.4 $<$ B-V $<$ 1.5} 
\end{array} 
\end{displaymath}

Here, ${\rm A}_{\rm B}$ and ${\rm A}_{\rm V}$ stand for the amplitudes in the B 
and 
V bands, respectively.
Regarding the color selection criterion, we note that the foreground reddening 
in the direction
to NGC 300 is only E(B-V)=0.02 (Burstein \& Heiles 1984) so that the observed 
colors correspond
closely to the intrinsic colors of the stars, supposing that intrinsic 
absorption
within NGC 300 is small. 

To derive the mean magnitudes for those variables passing the selection test for
a Cepheid, the light curves
were approximated by  Fourier series of an order ranging 
from 2 to 5, depending on the accuracy of the photometry and/or the phase 
coverage.  
Fig. 2 shows such fits for two exemplary B light curves:  one of a 
bright, well observed Cepheid, and one of a much fainter Cepheid having a 
noisier light curve
and less points because the star could only be detected and measured on nights 
of good
seeing. However, even in such cases mean magnitudes could be determined rather 
accurately.
They were obtained by fitting the light curves, converted to intensity 
units, and transforming the mean intensities back to a magnitude scale. 
The statistical accuracy of our intensity mean B and V magnitudes was typically 
better 
than 0.01 mag
(i.e. much smaller than the uncertainty of the zero point of our photometry).

Alltogether, 117 stars satisfied our selection criteria and entered our 
catalog of Cepheids presented in this paper. 
Table 2 contains their description. The first column is the star identification 
number. In the next columns the equatorial coordinates, derived periods, epochs 
of zero phase corresponding to maximum light in V, and intensity mean 
brightnesses 
in the B and V bands are given. The last column contains remarks on the objects. 
V2, V3, etc correspond
to the numbering system introduced by Graham (1984) for the 34 variable star
candidates in NGC 300 discovered by him. The digits of the periods given reflect 
their uncertainties.
The phased B and V light curves for all Cepheids are presented
in Fig. 3. Note that each observational point on the light curves is the 
mean from all observations (up to 5) obtained through a given filter during a 
particular night.

In addition to the 117 Cepheids in Table 2, we present another 12 Cepheid 
candidates in Table 3. 
These objects do not satisfy all criteria described above, but the shapes of 
their (noisy)
light curves suggest that they may be Cepheids, too. The phased B and V light 
curves of these objects are displayed in Fig. 4. For all Cepheids, the 
individual B and V 
observations are given in Table 4.

Finding charts for all Cepheid variables, prepared based on V band
image  are presented in Figure 5 (available only in the
electronic edition of the Journal). They can be also be obtained from the
authors upon request.

\begin{deluxetable}{c c c c c c c c}
\tablecaption{Cepheids in NGC 300}
\tablehead{
\colhead{ID} & \colhead{RA (J2000)} & \colhead{DEC (J2000)} &
\colhead{P} & \colhead{${\rm T}_{0}-2450000$} & \colhead{$<B>$} &
\colhead{$<V>$} &
\colhead{Remarks} \\
& & & [days] & [HJD] & [mag] & [mag] &
}
\startdata
cep001 & 0\uph55\upm11\zdot\ups60 & -37\arcd33\arcm54\zdot\arcs9 &  115    &  1487     &  21.14 &  20.08 & V24, blend \nl 
cep002 & 0\uph54\upm35\zdot\ups04 & -37\arcd35\arcm00\zdot\arcs6 &  89.05  &  1503.26  &  20.78 &  19.77 & V12, blend \nl 
cep003 & 0\uph54\upm54\zdot\ups32 & -37\arcd37\arcm01\zdot\arcs6 &  83     &  1404     &  20.13 &  19.26 & \nl 
cep004 & 0\uph54\upm53\zdot\ups98 & -37\arcd39\arcm30\zdot\arcs5 &  75     &  1394     &  20.69 &  19.78& V18, blend \nl 
cep005 & 0\uph54\upm08\zdot\ups90 & -37\arcd41\arcm40\zdot\arcs8 &  56.57  &  1411.68  &  21.12 &  20.39 & V3 \nl 
cep006 & 0\uph55\upm41\zdot\ups25 & -37\arcd49\arcm06\zdot\arcs5 &  52.754 &  1397.017 &  21.31 &  20.47 & V32 \nl 
cep007 & 0\uph54\upm23\zdot\ups79 & -37\arcd41\arcm04\zdot\arcs5 &  43.35  &  1425.47  &  21.86 &  20.87 & V8 \nl 
cep008 & 0\uph54\upm26\zdot\ups47 & -37\arcd42\arcm04\zdot\arcs8 &  40.6   &  1545.8   &  20.96 &  20.33 &   \nl 
cep009 & 0\uph54\upm41\zdot\ups97 & -37\arcd34\arcm44\zdot\arcs3 &  36.7   &  1520.1   &  21.97 &  21.06 & \nl 
cep010 & 0\uph54\upm48\zdot\ups50 & -37\arcd44\arcm58\zdot\arcs4 &  35.8   &  1531.3   &  22.44 &  21.28 & \nl 
cep011 & 0\uph54\upm57\zdot\ups23 & -37\arcd40\arcm26\zdot\arcs7 &  35.3   &  1487.8   &  21.98 &  21.17 & \nl 
cep012 & 0\uph55\upm16\zdot\ups16 & -37\arcd37\arcm26\zdot\arcs9 &  35.005 &  1436.958 &  21.61 &  20.86 & V27 \nl 
cep013 & 0\uph55\upm12\zdot\ups28 & -37\arcd42\arcm44\zdot\arcs0 &  34.72  &  1520.26  &  21.73 &  20.86 & V25 \nl 
cep014 & 0\uph54\upm41\zdot\ups73 & -37\arcd46\arcm15\zdot\arcs9 &  33.975 &  1405.393 &  21.43 &  20.69 & V13 \nl 
cep015 & 0\uph54\upm51\zdot\ups36 & -37\arcd39\arcm11\zdot\arcs0 &  32.3   &  1543.8   &  21.80 &  20.98 & \nl 
cep016 & 0\uph55\upm04\zdot\ups71 & -37\arcd43\arcm59\zdot\arcs0 &  28.6   &  1529.13  &  22.33 &  21.34 & \nl 
cep017 & 0\uph54\upm52\zdot\ups64 & -37\arcd35\arcm39\zdot\arcs7 &  28.5   &  1436.8   &  22.37 &  21.55 & \nl 
cep018 & 0\uph54\upm26\zdot\ups85 & -37\arcd34\arcm23\zdot\arcs9 &  25.010 &  1543.715 &  22.12 &  21.43 & V10 \nl 
cep019 & 0\uph55\upm25\zdot\ups94 & -37\arcd39\arcm42\zdot\arcs5 &  24.9   &  1528.6   &  22.48 &  21.60 & V28 \nl 
cep020 & 0\uph54\upm53\zdot\ups78 & -37\arcd41\arcm19\zdot\arcs2 &  24.4   &  1392.1   &  22.65 &  21.83 & \nl 
cep021 & 0\uph54\upm59\zdot\ups56 & -37\arcd41\arcm29\zdot\arcs9 &  24.228 &  1404.237 &  21.83 &  21.25 & V33 \nl 
cep022 & 0\uph55\upm36\zdot\ups56 & -37\arcd40\arcm09\zdot\arcs0 &  24.2   &  1496.7   &  22.36 &  21.64 & V31 \nl 
cep023 & 0\uph55\upm02\zdot\ups25 & -37\arcd45\arcm49\zdot\arcs2 &  24.033 &  1484.147 &  22.83 &  21.87 & V21 \nl 
cep024 & 0\uph54\upm45\zdot\ups52 & -37\arcd42\arcm18\zdot\arcs5 &  23.8   &  1496.4   &  21.87 &  21.12 & \nl 
cep025 & 0\uph54\upm25\zdot\ups97 & -37\arcd40\arcm38\zdot\arcs7 &  23.7   &  1391.7   &  22.49 &  21.64 & \nl 
cep026 & 0\uph55\upm33\zdot\ups14 & -37\arcd36\arcm17\zdot\arcs1 &  23.446 &  1408.859 &  21.85 &  21.21 & V29 \nl 
cep027 & 0\uph55\upm12\zdot\ups10 & -37\arcd37\arcm42\zdot\arcs5 &  23.3   &  1491.6   &  22.03 &  21.25 & \nl 
cep028 & 0\uph54\upm51\zdot\ups39 & -37\arcd39\arcm27\zdot\arcs0 &  23.1   &  1487.8   &  21.59 &  21.02 & \nl 
cep029 & 0\uph54\upm04\zdot\ups52 & -37\arcd39\arcm20\zdot\arcs5 &  22.6   &  1392.    &  22.64 &  21.83 & \nl 
cep030 & 0\uph54\upm46\zdot\ups63 & -37\arcd37\arcm14\zdot\arcs2 &  22.3   &  1517.5   &  22.35 &  21.53 & \nl 
cep031 & 0\uph55\upm32\zdot\ups91 & -37\arcd40\arcm45\zdot\arcs0 &  21.5   &  1407.5   &  23.03 &  21.95 & \nl 
cep032 & 0\uph54\upm59\zdot\ups48 & -37\arcd38\arcm34\zdot\arcs2 &  21.1   &  1496.6   &  22.09 &  21.41 & \nl 
cep033 & 0\uph55\upm07\zdot\ups02 & -37\arcd39\arcm40\zdot\arcs5 &  20.7   &  1401.0   &  22.32 &  21.74 & \nl 
cep034 & 0\uph55\upm02\zdot\ups54 & -37\arcd33\arcm18\zdot\arcs8 &  20.64  &  1390.76  &  22.50 &  21.78 & V22 \nl 
cep035 & 0\uph54\upm44\zdot\ups65 & -37\arcd38\arcm30\zdot\arcs6 &  19.4   &  1519.3   &  22.35 &  21.46 & \nl 
cep036 & 0\uph55\upm08\zdot\ups33 & -37\arcd38\arcm11\zdot\arcs8 &  18.9   &  1488.6   &  23.25 &  22.47 & \nl 
cep037 & 0\uph54\upm37\zdot\ups07 & -37\arcd39\arcm05\zdot\arcs5 &  18.7   &  1545.6   &  23.37 &  22.43 & \nl 
cep038 & 0\uph55\upm31\zdot\ups34 & -37\arcd34\arcm31\zdot\arcs4 &  18.2   &  1408.9   &  22.12 &  21.53 & \nl 
cep039 & 0\uph55\upm04\zdot\ups12 & -37\arcd46\arcm14\zdot\arcs1 &  18.2   &  1391.4   &  23.05 &  22.13 & \nl 
cep040 & 0\uph54\upm25\zdot\ups00 & -37\arcd37\arcm57\zdot\arcs7 &  18.213 &  1545.589 &  22.00 &  21.49 & V9 \nl 
cep041 & 0\uph55\upm01\zdot\ups94 & -37\arcd45\arcm45\zdot\arcs0 &  18.012 &  1392.004 &  22.00 &  21.43 & V20 \nl 
cep042 & 0\uph53\upm37\zdot\ups17 & -37\arcd26\arcm00\zdot\arcs6 &  17.93  &  1408.87  &  22.88 &  22.02 & \nl 
cep043 & 0\uph54\upm02\zdot\ups20 & -37\arcd39\arcm02\zdot\arcs2 &  17.833 &  1524.816 &  22.07 &  21.51 & V2 \nl 
cep044 & 0\uph54\upm59\zdot\ups73 & -37\arcd43\arcm03\zdot\arcs0 &  17.25  &  1408.30  &  22.89 &  22.06 & \nl 
cep045 & 0\uph55\upm29\zdot\ups33 & -37\arcd41\arcm23\zdot\arcs0 &  17.0   &  1548.0   &  22.16 &  21.66 & \nl 
cep046 & 0\uph55\upm21\zdot\ups40 & -37\arcd33\arcm48\zdot\arcs3 &  16.5   &  1392.9   &  23.34 &  22.47 & \nl 
cep047 & 0\uph54\upm39\zdot\ups67 & -37\arcd42\arcm38\zdot\arcs7 &  16.5   &  1496.6   &  23.44 &  22.21 & \nl 
cep048 & 0\uph55\upm35\zdot\ups68 & -37\arcd44\arcm12\zdot\arcs1 &  16.5   &  1390.6   &  22.12 &  21.51 & \nl 
cep049 & 0\uph55\upm14\zdot\ups64 & -37\arcd43\arcm37\zdot\arcs6 &  16.2   &  1390.8   &  23.00 &  22.07 & \nl 
cep050 & 0\uph54\upm30\zdot\ups22 & -37\arcd43\arcm04\zdot\arcs4 &  15.9   &  1438.2   &  22.39 &  21.74 & \nl 
cep051 & 0\uph54\upm22\zdot\ups63 & -37\arcd34\arcm48\zdot\arcs0 &  15.72  &  1543.42  &  22.64 &  22.00 & \nl 
cep052 & 0\uph55\upm22\zdot\ups01 & -37\arcd34\arcm22\zdot\arcs6 &  15.6   &  1498.8   &  22.22 &  21.55 & V26 \nl 
cep053 & 0\uph55\upm20\zdot\ups84 & -37\arcd43\arcm03\zdot\arcs0 &  15.5   &  1520.5   &  23.11 &  22.21 & \nl 
cep054 & 0\uph54\upm09\zdot\ups33 & -37\arcd40\arcm51\zdot\arcs8 &  15.4   &  1481.4   &  22.21 &  21.70 & V4 \nl 
cep055 & 0\uph55\upm06\zdot\ups92 & -37\arcd44\arcm36\zdot\arcs4 &  15.0   &  1404.9   &  22.24 &  21.66 & \nl 
cep056 & 0\uph54\upm37\zdot\ups36 & -37\arcd38\arcm05\zdot\arcs8 &  15.05  &  1431.95  &  22.77 &  22.22 & \nl 
cep057 & 0\uph55\upm08\zdot\ups19 & -37\arcd35\arcm09\zdot\arcs6 &  14.8   &  1485.3   &  22.91 &  22.17 & \nl 
cep058 & 0\uph55\upm02\zdot\ups26 & -37\arcd47\arcm00\zdot\arcs6 &  14.80  &  1519.48  &  23.00 &  22.14 & \nl 
cep059 & 0\uph54\upm48\zdot\ups55 & -37\arcd36\arcm43\zdot\arcs0 &  14.5   &  1392.7   &  22.97 &  22.10 & \nl 
cep060 & 0\uph55\upm06\zdot\ups98 & -37\arcd41\arcm32\zdot\arcs9 &  14.40  &  1425.23  &  23.16 &  22.20 & \nl 
cep061 & 0\uph55\upm18\zdot\ups96 & -37\arcd39\arcm11\zdot\arcs3 &  14.3   &  1519.4   &  22.99 &  22.24 & \nl 
cep062 & 0\uph55\upm06\zdot\ups82 & -37\arcd35\arcm13\zdot\arcs4 &  14.3   &  1436.6   &  22.76 &  22.09 & \nl 
cep063 & 0\uph54\upm12\zdot\ups35 & -37\arcd39\arcm19\zdot\arcs1 &  14.3   &  1392.8   &  22.64 &  21.98 & V7 \nl 
cep064 & 0\uph54\upm53\zdot\ups97 & -37\arcd44\arcm42\zdot\arcs5 &  14.3   &  1392.5   &  23.62 &  22.33 & \nl 
cep065 & 0\uph55\upm20\zdot\ups28 & -37\arcd39\arcm38\zdot\arcs6 &  14.1   &  1424.9   &  22.45 &  21.81 & \nl 
cep066 & 0\uph55\upm29\zdot\ups53 & -37\arcd42\arcm08\zdot\arcs4 &  13.8   &  1392.7   &  22.77 &  22.10 & \nl 
cep067 & 0\uph54\upm20\zdot\ups01 & -37\arcd38\arcm32\zdot\arcs8 &  13.8   &  1491.5   &  23.32 &  22.45 & \nl 
cep068 & 0\uph54\upm29\zdot\ups75 & -37\arcd40\arcm46\zdot\arcs1 &  13.8   &  1437.5   &  23.10 &  22.49 & \nl 
cep069 & 0\uph54\upm48\zdot\ups86 & -37\arcd37\arcm21\zdot\arcs1 &  13.6   &  1496.6   &  22.93 &  22.30 & \nl 
cep070 & 0\uph54\upm41\zdot\ups40 & -37\arcd33\arcm48\zdot\arcs1 &  13.57  &  1408.64  &  22.57 &  22.05 & \nl 
cep071 & 0\uph54\upm37\zdot\ups35 & -37\arcd37\arcm32\zdot\arcs4 &  13.54  &  1438.83  &  22.49 &  21.87 & \nl 
cep072 & 0\uph54\upm49\zdot\ups87 & -37\arcd36\arcm57\zdot\arcs7 &  13.5   &  1547.9   &  23.23 &  22.33 & \nl 
cep073 & 0\uph54\upm34\zdot\ups77 & -37\arcd39\arcm53\zdot\arcs9 &  13.4   &  1543.7   &  22.59 &  21.97 & \nl 
cep074 & 0\uph54\upm23\zdot\ups07 & -37\arcd35\arcm42\zdot\arcs2 &  13.3   &  1438.7   &  22.52 &  21.89 & \nl 
cep075 & 0\uph54\upm14\zdot\ups77 & -37\arcd25\arcm48\zdot\arcs2 &  13.3   &  1526.1   &  22.77 &  22.13 & blend \nl 
cep076 & 0\uph54\upm19\zdot\ups58 & -37\arcd30\arcm24\zdot\arcs4 &  13.1   &  1519.6   &  22.53 &  21.86 & \nl 
cep077 & 0\uph55\upm15\zdot\ups09 & -37\arcd34\arcm33\zdot\arcs4 &  13.0   &  1484.5   &  22.78 &  22.21 & \nl 
cep078 & 0\uph54\upm26\zdot\ups76 & -37\arcd44\arcm12\zdot\arcs5 &  12.5   &  1436.8   &  23.04 &  22.47 & \nl 
cep079 & 0\uph55\upm07\zdot\ups32 & -37\arcd40\arcm01\zdot\arcs8 &  11.9   &  1518.5   &  22.56 &  22.06 & \nl 
cep080 & 0\uph54\upm59\zdot\ups33 & -37\arcd44\arcm25\zdot\arcs6 &  11.8   &  1431.8   &  22.34 &  21.70 & \nl 
cep081 & 0\uph54\upm19\zdot\ups32 & -37\arcd44\arcm54\zdot\arcs1 &  11.6   &  1481.6   &  22.86 &  22.25 & \nl 
cep082 & 0\uph55\upm01\zdot\ups56 & -37\arcd35\arcm43\zdot\arcs6 &  11.5   &  1516.9   &  23.26 &  22.57 & \nl 
cep083 & 0\uph54\upm59\zdot\ups60 & -37\arcd47\arcm31\zdot\arcs7 &  11.4   &  1425.1   &  22.47 &  21.92 & blend \nl 
cep084 & 0\uph55\upm10\zdot\ups22 & -37\arcd44\arcm05\zdot\arcs8 &  11.3   &  1496.7   &  22.77 &  21.97 & \nl 
cep085 & 0\uph54\upm19\zdot\ups73 & -37\arcd37\arcm47\zdot\arcs7 &  11.2   &  1496.0   &  22.60 &  22.21 & \nl 
cep086 & 0\uph55\upm12\zdot\ups49 & -37\arcd36\arcm11\zdot\arcs3 &  10.6   &  1409.3   &  23.35 &  22.61 & \nl 
cep087 & 0\uph54\upm55\zdot\ups63 & -37\arcd32\arcm20\zdot\arcs2 &   9.5   &  1496.6   &  22.96 &  22.45 & \nl 
cep088 & 0\uph55\upm29\zdot\ups72 & -37\arcd46\arcm18\zdot\arcs6 &   9.38  &  1484.53  &  23.26 &  22.51 & \nl 
cep089 & 0\uph54\upm23\zdot\ups59 & -37\arcd35\arcm46\zdot\arcs4 &   9.37  &  1521.73  &  22.82 &  22.16 & \nl 
cep090 & 0\uph55\upm22\zdot\ups61 & -37\arcd34\arcm23\zdot\arcs7 &   9.3   &  1491.8   &  22.71 &  22.33 & \nl 
cep091 & 0\uph54\upm24\zdot\ups31 & -37\arcd35\arcm46\zdot\arcs0 &   9.2   &  1481.8   &  23.26 &  22.62 & \nl 
cep092 & 0\uph55\upm14\zdot\ups60 & -37\arcd36\arcm02\zdot\arcs7 &   9.07  &  1491.47  &  23.07 &  22.55 & \nl 
cep093 & 0\uph55\upm01\zdot\ups83 & -37\arcd37\arcm52\zdot\arcs1 &   9.06  &  1405.52  &  23.02 &  22.58 & \nl 
cep094 & 0\uph54\upm50\zdot\ups87 & -37\arcd35\arcm49\zdot\arcs1 &   8.89  &  1400.93  &  22.92 &  22.23 & blend \nl 
cep095 & 0\uph55\upm01\zdot\ups43 & -37\arcd37\arcm42\zdot\arcs5 &   8.85  &  1547.45  &  22.93 &  22.11 & \nl 
cep096 & 0\uph55\upm54\zdot\ups92 & -37\arcd46\arcm39\zdot\arcs2 &   8.76  &  1543.29  &  23.01 &  22.36 & \nl 
cep097 & 0\uph54\upm25\zdot\ups36 & -37\arcd31\arcm58\zdot\arcs7 &   8.46  &  1424.44  &  22.89 &  22.36 & \nl 
cep098 & 0\uph55\upm40\zdot\ups50 & -37\arcd37\arcm47\zdot\arcs6 &   8.42  &  1491.66  &  23.07 &  22.50 & \nl 
cep099 & 0\uph55\upm34\zdot\ups52 & -37\arcd42\arcm42\zdot\arcs4 &   8.30  &  1436.72  &  23.04 &  22.48 & \nl 
cep100 & 0\uph55\upm17\zdot\ups53 & -37\arcd37\arcm24\zdot\arcs2 &   8.07  &  1393.79  &  22.71 &  22.21 & \nl 
cep101 & 0\uph54\upm40\zdot\ups35 & -37\arcd33\arcm57\zdot\arcs8 &   8.0   &  1481.5   &  22.53 &  21.95 & \nl 
cep102 & 0\uph54\upm29\zdot\ups74 & -37\arcd39\arcm04\zdot\arcs0 &   7.9   &  1526.3   &  23.28 &  22.80 & \nl 
cep103 & 0\uph54\upm35\zdot\ups68 & -37\arcd44\arcm54\zdot\arcs2 &   7.85  &  1524.82  &  23.32 &  22.78 & \nl 
cep104 & 0\uph54\upm39\zdot\ups54 & -37\arcd35\arcm28\zdot\arcs0 &   7.75  &  1409.22  &  23.34 &  22.77 & \nl 
cep105 & 0\uph54\upm30\zdot\ups73 & -37\arcd41\arcm50\zdot\arcs6 &   7.75  &  1524.68  &  23.11 &  22.61 & \nl 
cep106 & 0\uph54\upm18\zdot\ups66 & -37\arcd45\arcm59\zdot\arcs2 &   7.67  &  1400.70  &  23.06 &  22.49 & \nl 
cep107 & 0\uph54\upm41\zdot\ups41 & -37\arcd44\arcm02\zdot\arcs8 &   7.65  &  1517.17  &  23.17 &  22.58 & \nl 
cep108 & 0\uph55\upm28\zdot\ups33 & -37\arcd34\arcm40\zdot\arcs7 &   7.61  &  1526.54  &  23.08 &  22.68 & \nl 
cep109 & 0\uph54\upm06\zdot\ups59 & -37\arcd35\arcm43\zdot\arcs3 &   7.38  &  1516.83  &  23.11 &  22.56 & \nl 
cep110 & 0\uph54\upm39\zdot\ups57 & -37\arcd41\arcm17\zdot\arcs6 &   7.15  &  1521.79  &  23.01 &  22.56 & blend \nl 
cep111 & 0\uph54\upm07\zdot\ups72 & -37\arcd39\arcm51\zdot\arcs9 &   7.10  &  1521.52  &  22.99 &  22.59 & \nl 
cep112 & 0\uph55\upm01\zdot\ups03 & -37\arcd45\arcm25\zdot\arcs2 &   6.92  &  1545.83  &  23.43 &  22.93 & \nl 
cep113 & 0\uph55\upm05\zdot\ups93 & -37\arcd37\arcm13\zdot\arcs1 &   6.6   &  1526.8   &  22.97 &  22.40 & \nl 
cep114 & 0\uph55\upm33\zdot\ups15 & -37\arcd37\arcm01\zdot\arcs8 &   6.49  &  1424.79  &  23.13 &  22.73 & \nl 
cep115 & 0\uph55\upm01\zdot\ups70 & -37\arcd39\arcm02\zdot\arcs1 &   6.25  &  1521.65  &  23.08 &  22.57 & \nl 
cep116 & 0\uph54\upm53\zdot\ups41 & -37\arcd35\arcm59\zdot\arcs1 &   5.98  &  1521.71  &  23.79 &  23.08 & \nl 
cep117 & 0\uph54\upm53\zdot\ups53 & -37\arcd42\arcm25\zdot\arcs6 &   5.42  &  1433.79  &  23.02 &  22.48 & \nl 
\enddata

\end{deluxetable}

\begin{deluxetable}{c c c c c c c c}
\tablecaption{Cepheid Candidates in NGC 300}
\tablehead{
\colhead{ID} & \colhead{RA (J2000)} & \colhead{DEC (J2000)} &
\colhead{P} & \colhead{${\rm T}_{0}-2450000$} & \colhead{$<B>$} &
\colhead{$<V>$} &
\colhead{Remarks}\\
& & & [days] & [HJD] & [mag] & [mag] &
}
\startdata

cep118 & 0\uph54\upm40\zdot\ups06 & -37\arcd37\arcm53\zdot\arcs1 &  13.8   &  1525.1   &  22.76 &  22.11 & \nl 
cep119 & 0\uph54\upm16\zdot\ups94 & -37\arcd37\arcm23\zdot\arcs5 &  13.4   &  1482.8   &  22.81 &  22.12 & \nl 
cep120 & 0\uph54\upm20\zdot\ups33 & -37\arcd34\arcm24\zdot\arcs1 &  11.82  &  1430.98  &  22.76 &  22.64 & blend \nl 
cep121 & 0\uph54\upm32\zdot\ups70 & -37\arcd37\arcm47\zdot\arcs8 &  11.3   &  1526.6   &  23.14 &  22.37 & \nl 
cep122 & 0\uph54\upm59\zdot\ups91 & -37\arcd36\arcm27\zdot\arcs9 &  10.5   &  1517.1   &  22.77 &  22.39 & blend \nl 
cep123 & 0\uph55\upm34\zdot\ups95 & -37\arcd41\arcm22\zdot\arcs4 &   8.7   &  1492.0   &  22.75 &  22.20 & blend \nl 
cep124 & 0\uph54\upm50\zdot\ups09 & -37\arcd42\arcm18\zdot\arcs3 &   8.3   &  1433.8   &  22.95 &  22.38 & blend \nl 
cep125 & 0\uph54\upm26\zdot\ups88 & -37\arcd38\arcm39\zdot\arcs7 &   7.97  &  1524.61  &  23.05 &  22.72 & \nl 
cep126 & 0\uph55\upm22\zdot\ups91 & -37\arcd43\arcm48\zdot\arcs3 &   7.40  &  1488.55  &  23.14 &  22.72 & \nl 
cep127 & 0\uph54\upm20\zdot\ups58 & -37\arcd34\arcm45\zdot\arcs3 &   7.30  &  1431.85  &  23.12 &  22.82 & \nl 
cep128 & 0\uph55\upm33\zdot\ups81 & -37\arcd42\arcm39\zdot\arcs4 &   6.15  &  1545.47  &  23.13 &  22.81 & \nl 
cep129 & 0\uph54\upm49\zdot\ups47 & -37\arcd43\arcm58\zdot\arcs8 &   5.37  &  1516.64  &  22.83 &  22.47 & \nl 
\enddata

\end{deluxetable}

\begin{deluxetable}{ccccc}
\tablecaption{Individual B and V Observations}
\tablehead{
\colhead{object}  & \colhead{filter} &
\colhead{HJD-2450000}  & \colhead{mag}  & \colhead{$\sigma_{mag}$}\\
}
\startdata
cep001 &  B & 1390.790 & 20.786 &  0.027\\
cep001 &  B & 1391.939 & 20.826 &  0.011\\
cep001 &  B & 1392.777 & 20.839 &  0.027\\
cep001 &  B & 1393.831 & 20.812 &  0.029\\
cep001 &  B & 1400.918 & 20.882 &  0.011\\
cep001 &  B & 1404.815 & 20.979 &  0.013\\
cep001 &  B & 1408.822 & 21.039 &  0.010\\
cep001 &  B & 1424.819 & 21.455 &  0.038\\
cep001 &  B & 1431.808 & 21.636 &  0.011\\
cep001 &  B & 1433.803 & 21.584 &  0.021\\
cep001 &  B & 1436.809 & 21.705 &  0.015\\
cep001 &  B & 1438.815 & 21.768 &  0.020\\
cep001 &  B & 1481.586 & 20.342 &  0.017\\
cep001 &  B & 1484.586 & 20.105 &  0.013\\
cep001 &  B & 1488.546 & 20.162 &  0.013\\
cep001 &  B & 1491.565 & 20.241 &  0.011\\
cep001 &  B & 1496.517 & 20.414 &  0.029\\
cep001 &  B & 1498.566 & 20.511 &  0.014\\
cep001 &  B & 1516.594 & 20.938 &  0.010\\
cep001 &  B & 1519.570 & 21.004 &  0.006\\
cep001 &  B & 1521.572 & 21.051 &  0.009\\
cep001 &  B & 1524.577 & 21.081 &  0.010\\
cep001 &  B & 1526.549 & 21.197 &  0.022\\
cep001 &  B & 1528.613 & 21.104 &  0.038\\
cep001 &  B & 1531.572 & 21.086 &  0.049\\
cep001 &  B & 1543.549 & 21.514 &  0.027\\
cep001 &  B & 1545.636 & 21.469 &  0.027\\
cep001 &  B & 1547.560 & 21.596 &  0.027\\
\enddata
\tablecomments{The complete version of this table is in the electronic
edition of the Journal.  The printed edition contains only 
the B data for the Cepheid variable cep001.}
\end{deluxetable}

\section{Previously Known Cepheids}
As mentioned before, a previous survey for variable stars in NGC 300 has already 
been conducted
by Graham (1984). His observations, 
based on photographic plates taken at the prime focus of the CTIO 4 m telescope 
between November 1974
and October 1981, yielded a list 
of 34  variables or variable star candidates which included 18 Cepheids 
with derived periods,
and another 5 objects suspected to be Cepheids, but without enough data for a 
period determination.
Later Freedman et al. (1992) obtained CCD BVRI photometry at the same telescope 
for 16 
of the Cepheids discovered by Graham . The baseline of their new observations 
was about 
3 years, and as expected the quality of the CCD data was much higher than the 
previous photographic observations.
 However, in many cases the phase coverage of the resulting light curves was 
 poor,  making the determination of the mean 
brightnesses rather uncertain. Combining the photographic and their own CCD 
data, Freedman et al. 
(1992) had already obtained refined periods for some of the Cepheids. 

From our new WFI data we rediscovered all 18 Cepheids reported by 
Graham (1984) (see Table 2). We also confirm 3 of his Cepheid candidates to be 
truly Cepheids
(V7, V20, and V31). All the Graham variables were among the brightest objects in 
our database and have the most 
accurate photometry and the best coverage of their light curves.
We also found that the object designated 
by Graham as V4 and classified by him as an eclipsing binary is
another Cepheid with 
P = 15.37 d. Most probably the spurious classification was due to poor 
resolution of the
photographic data. We did not confirm the variability of the objects marked as 
V14 and V19. Regarding the periods, our values generally agree, within the 
quoted uncertainties, with 
those obtained in the past. However, in the case of 5 variables (V4, V18, V21, 
V24, V25) we 
found a significantly 
different value for the period. The discrepancy in the period determination 
for V24 may be the result of a genuine period change for this variable during 
the 
last 26 years. Unfortunately the previous observations are too scarce and 
uncertain to confirm this beyond doubt. For V21 we obtained a period 
of about 24.0 days instead of 9.667 days derived previously by Freedman et al.
However, the Freedman et al. observations for V21 do fit much better with our 
period than with that derived by themselves. The same happens in the case of 
star V25. 
For both of these Cepheids the previous data look much better when phased with 
our periods
than with those obtained before, and  we were able to refine the periods 
(see below) combining our new with the previous data. The photographic data 
obtained for 
stars V4 and V18 are not accurate enough to be useful to improve the periods 
from the current new data. 

Including our new observations, the time baseline since the first epoch 
observations is now 26 years, corresponding 
to about 76-1000 pulsation cycles for the Cepheids, which makes further 
improvement of the periods of 
these stars possible. 
Before doing this we need to make sure that all the data sets share the same 
photometric zero point. Walker (1988) had already shown and corrected a zero 
point and scale error in 
Graham's photometric data. Using this correction, we tied the Graham data to 
Freedman's CCD system 
using the relations derived by Freedman et al. (1992). Finally, we need to check
if the Freedman et al. (1992) photometry is not significantly different from 
ours. We did this by comparing the intensity mean magnitudes derived from our 
measurements
to those given by Freedman et al., for the common stars. Fig. 6
presents this comparison for 8 Cepheids with reasonably
good photometry from Freedman et al. It is seen that both data sets yield mean 
magnitudes which
 are in very good agreement. We also checked the agreement 
between individual observations from these two data sets. An example is given in 
Fig. 7 for the variable V3.
 It is seen that a systematic zero point offset, if any, is very small, and that 
the individual observations
 generally agree very well. It is also worth noting that apart from their much 
larger scatter, the photographic 
data fit also quite well  our observations.

After making sure that there are no significant shifts in photometric zero 
points
of the three available data sets, we phased all the data for common objects 
with the best available period (usually that from Freedman et al., (1992)), 
looking 
for phase shifts to our present new data. In 15 cases we found clear shifts in 
the phase indicating 
the need to adjust the period. The periods of these stars given in Table 1 are 
already 
the improved periods resulting from a combination of the available data sets. 
For 7 remaining Cepheids, the long gaps between the individual 
data sets, the  accuracy of the available periods, the large  scatter in the 
photographic data, and a very limited number of  previous CCD observations (if 
available)
 did not allow us to refine the periods that we derived from our data alone.

\section{Summary and Discussion}
Results are presented of a wide-field multi-color photometric monitoring of NGC 
300. Based on
 observations obtained on 29 nights during a 5.3 month period, 
117 Cepheids and 12 Cepheid candidates were discovered from images in B and V 
bands. Their periods range 
between 115  and 5.42 days.
For each object we provide equatorial coordinates, period, intensity mean 
B and V magnitudes and phased light curves. The accuracy of the derived periods 
is generally about $0.5-1 * 10^{-2}*$ P. We have re-discovered all previously 
known Cepheid variables,
and for many of these the periods could be improved by combining the available 
data sets.
 The photometric zero points in both B and V are secure to
within 0.03 mag.

The spatial distribution 
of the detected Cepheids is shown in Fig. 1. It can be appreciated that we were 
able to detect 
Cepheids even in the very dense regions near the center of NGC 300. 
Since the Cepheid distribution traces the spiral arms of the galaxy, most of our 
Cepheids
lie in dense fields and are most probably blended, to some degree, with 
unresolved stars. This
blending problem will be discussed in more detail in a forthcoming paper in 
which we will
derive the distance of NGC 300 based on Cepheid PL relations.     

In Fig. 8, we show the location of the detected Cepheids on the observed V, B-V 
color-magnitude diagram 
constructed from our data. The Cepheids delineate the typical Cepheid 
instability strip. Fig. 7 also
 indicates that the completeness of our Cepheid search drops suddenly for 
objects 
fainter than about 22.5 mag in V. For magnitudes brighter than this, we expect 
the 
survey
to be near-complete. The corresponding period cutoff is close to 10 days, in the 
sense that our
survey should have discovered a very large fraction of the Cepheids in NGC 300 
with periods
larger than 10 days.

Our Cepheid catalog contains 22 entries common to the list of variable stars 
prepared 
by Graham (1984). Sixteen of the Cepheids had also accurate (albeit sometimes 
sparse)
 previous CCD photometry (Freedman et al. 
1992). We found very good agreement between the zero points in B and V of our 
data and those 
presented by Freedman et al. Using combined data sets, the periods of common 
objects 
were examined and improved whenever possible. 

Fig. 9 presents the period distribution for all the 117 Cepheids in our catalog. 
There are three evident peaks, corresponding to periods of about 8, 14 and 24 
days. 
Their existence and locations do not depend on the size 
of the binning we are using. A precise interpretation of the physical meaning 
of these peaks is difficult (and beyond the scope of this paper) but one 
possibility 
is that they reflect the fact that the Cepheids we are observing in NGC 300 have 
been formed during different bursts of star formation, in different places in 
the 
galaxy.

Finally, we show the period-luminosity relations in B and V derived from our 
data 
in Fig. 10. For periods below about 10 days, the data are clearly affected by a 
Malmquist bias-only
the brighter Cepheids, at any given period below 10 days, have been detected in 
our survey. 
A detailed analysis of the PL relations to determine the distance of NGC 300 
will be the
subject of a forthcoming paper, which will include the presentation and analysis 
of
photometric data obtained at longer wavelengths.

\acknowledgments
We are very grateful to the European Southern Observatory for providing large 
amounts 
of observing time with the Wide Field Camera on the 2.2 m ESO/MPI telescope to 
this survey. It is 
a great pleasure to thank the 2p2 team for expert support. Among the astronomers 
who 
have been especially helpful with
the observations we would like to thank H. Boehnhardt, P. Leisy, J. Brewer, E. 
Pompei, D. Bersier, H. Jones, P. Francois and T. Augusteijn.
Pierre Leisy and L. Rizzi were very helpful with the data reductions.
Furthermore, we would like to thank Drs Andrzej Udalski, Micha{\l} 
Szyma{\'n}ski, 
and Grzegorz Pojma{\'n}ski for making their computer programs available and 
valuable discussion. WG 
would like to acknowledge financial
support from Fondecyt grant 8000002, and express his thanks to the 
Centrum fuer internationale
Migration und Entwicklung (CIM) in Germany for providing funds for a Sun Ultra 
60 workstation on
which most of the present data analysis was carried out.
We also acknowledge use of the Digitized Sky Survey which was produced at the 
Space Telescope Science Institute based on photographic data obtained with 
the UK Schmidt Telescope, operated by the Royal Observatory Edinburgh.

\begin{figure}[htb]
\vspace*{1 cm}
\caption{DSS map of NGC300. Red and yellow points corespond to the positions of 
detected Cepheids with periods smaller and larger than 15 days, respectively.
North is up and east to the left. The field of view is about 34 x 34 arcmin.
This figure is too big to present it here.}
\end{figure}

\begin{figure}[htb]
\vspace*{6cm}
\includegraphics{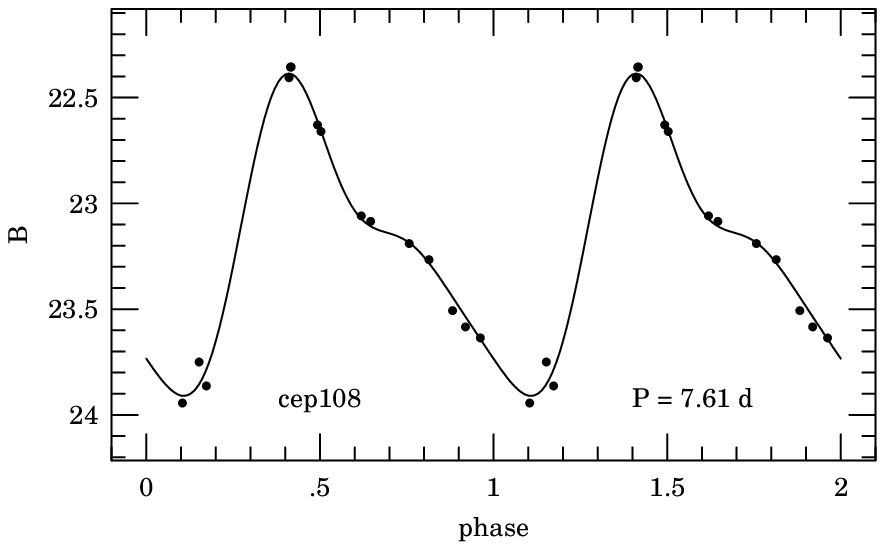}
\includegraphics{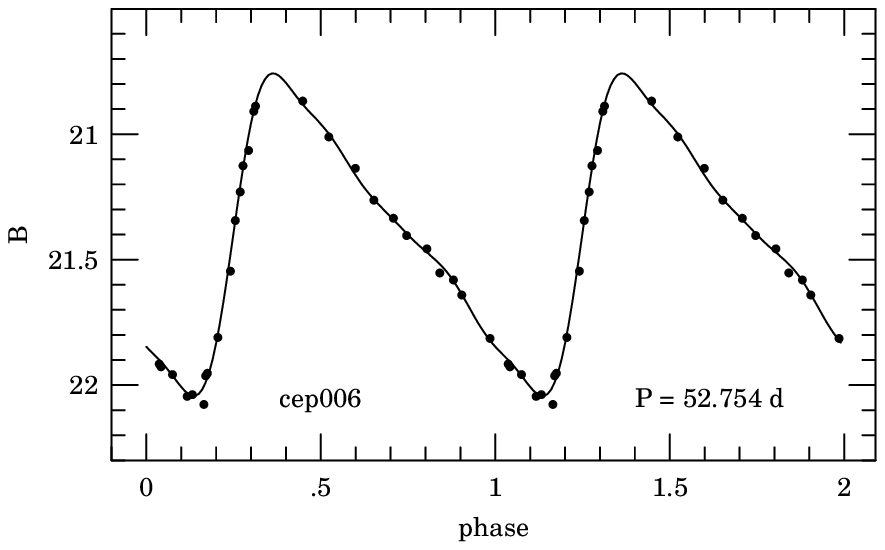}
\caption{Two exemplary B light curves of Cepheid variables in NGC 300, with 
fitted Fourier series of 
order 3 (left panel), and 5 (right panel). Phase points are repeated for 
clarity. It is seen that
good fits can be derived even for faint Cepheids with relatively noisy light 
curves.}
\end{figure}

\begin{figure}[htb]
\vspace*{20 cm}
\includegraphics{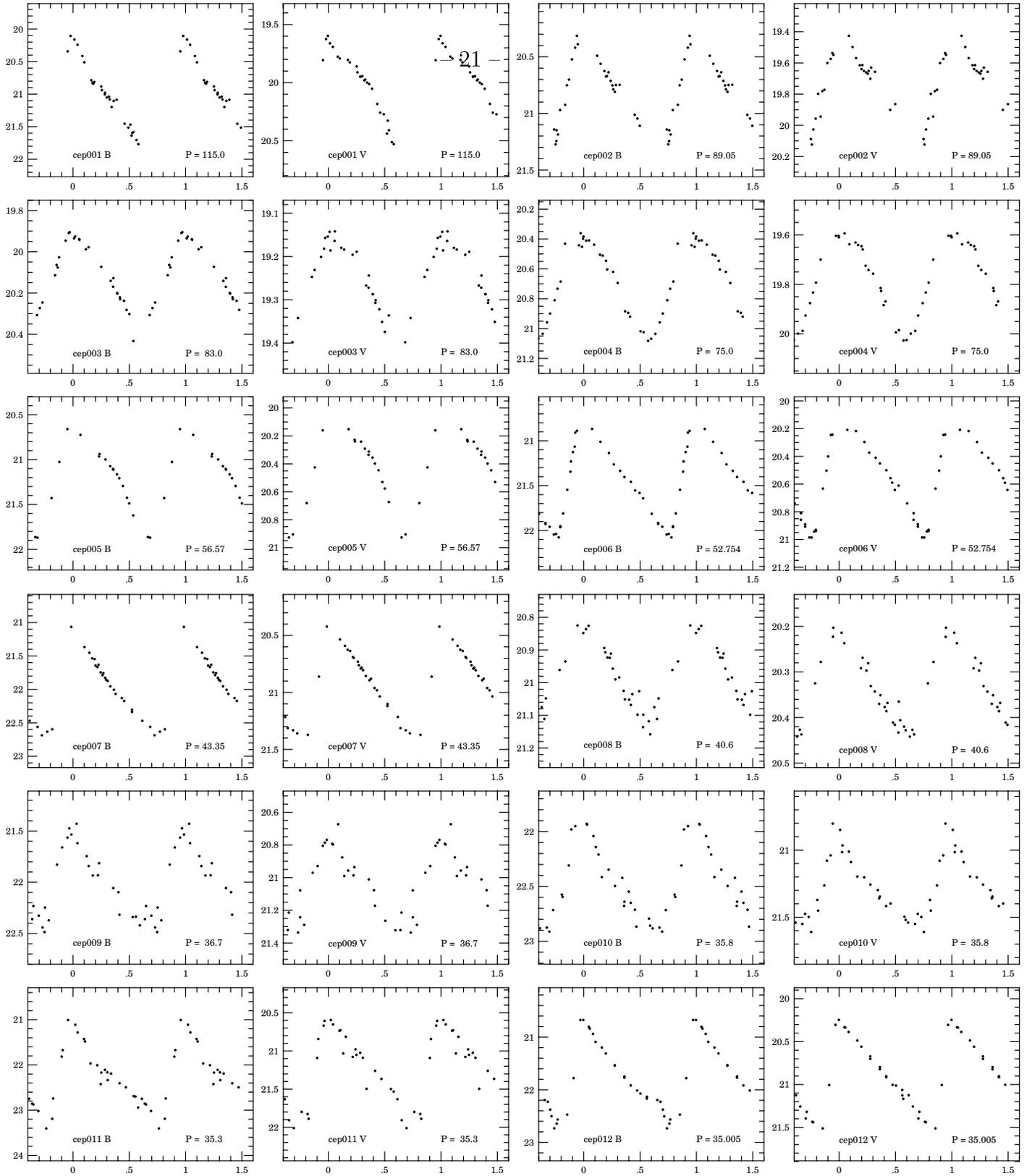}
\caption{Sample of light curves in B and V bands for  Cepheids in NGC 300 detected in 
our survey. Each point 
corresponds to a mean from  3-5 consecutive 360 s exposures from a given night.
}
\end{figure}

\begin{figure}[htb]
\caption{Light curves of additional Cepheid candidates in NGC 300. 
Not included here.}
\end{figure}

\begin{figure}[htb]
\caption{Finding charts for Cepheids and Cepheid candidates (not included here). 
The size of each map  is about 0.5 x 0.5 arcmin. North is up and East to the left. }   
\end{figure}

\begin{figure}[htb]
\vspace*{12cm}
\includegraphics{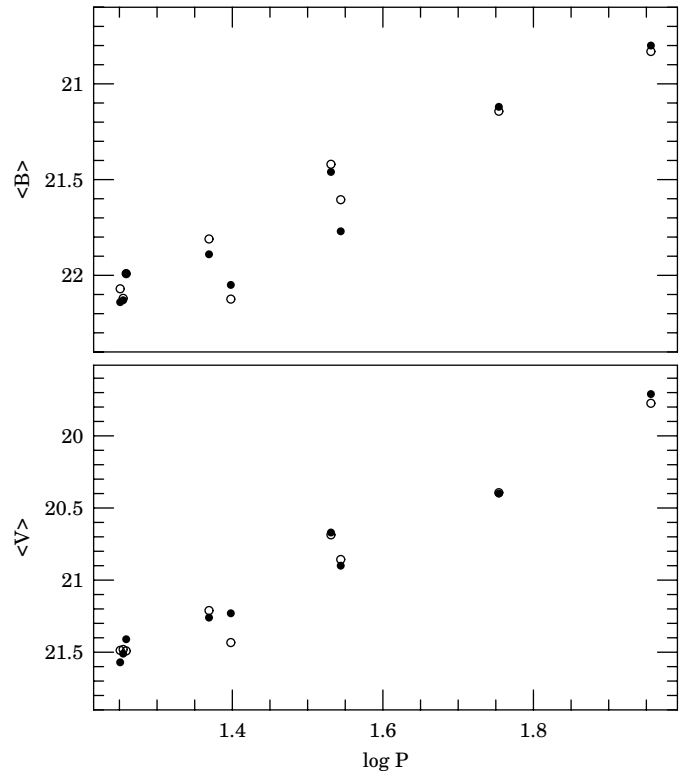}
\caption{Mean B and V brightness versus logarithm of period for 8
Cepheids common to our catalog and Freedman et al. (1992). Filled and open 
circles correspond to our, and Freedman et al. results, respectively.}
\end{figure}

\begin{figure}[htb]
\vspace*{6 cm}  
\includegraphics{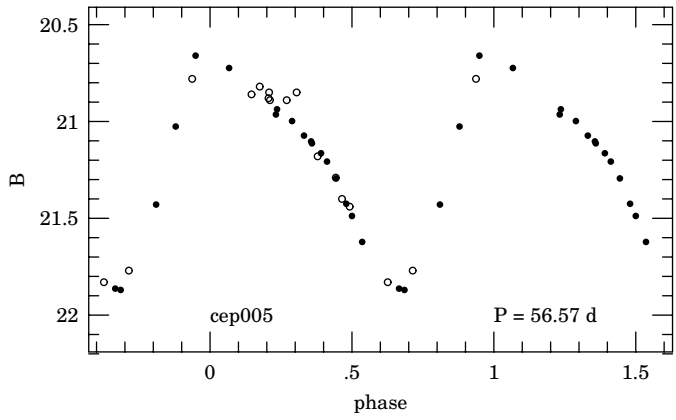}
\caption{Combined B light curve for the Cepheid cep005. Filled and open circles 
correspond to our, and Freedman et al. data, respectively. The good agreement
between the two data sets is demonstrated.}
\end{figure}

\begin{figure}[htb]
\vspace*{15cm}
\includegraphics{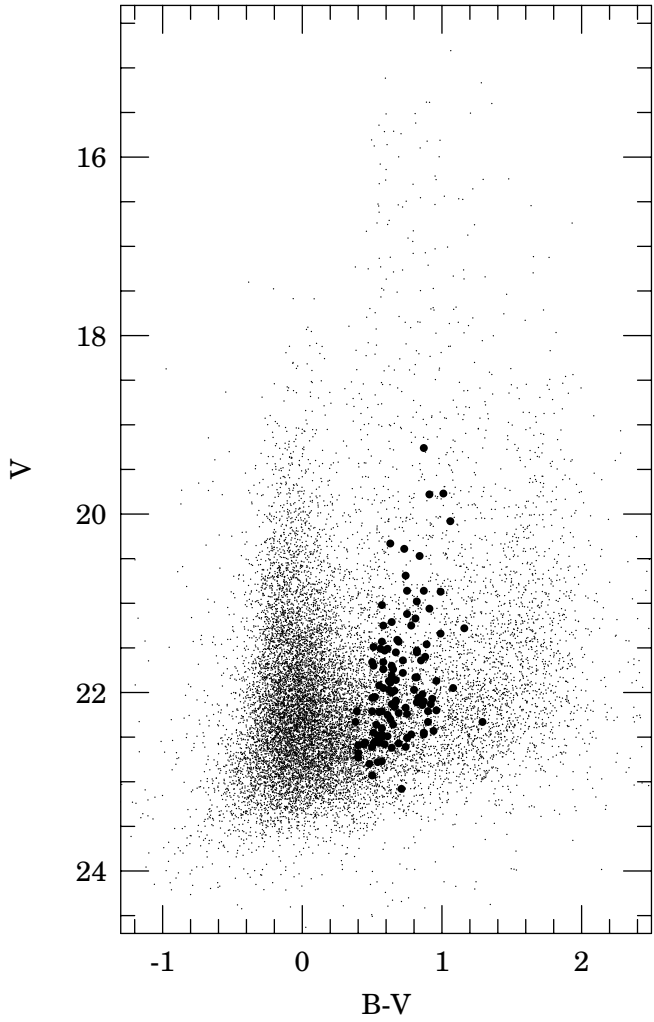}
\caption{V, B-V color-magnitude diagram for stars in NGC 300 with photometry 
from our WFI data.
The positions of the discovered Cepheids are marked with the filled circles.}
\end{figure}

\begin{figure}[htb]
\vspace*{6cm}
\includegraphics{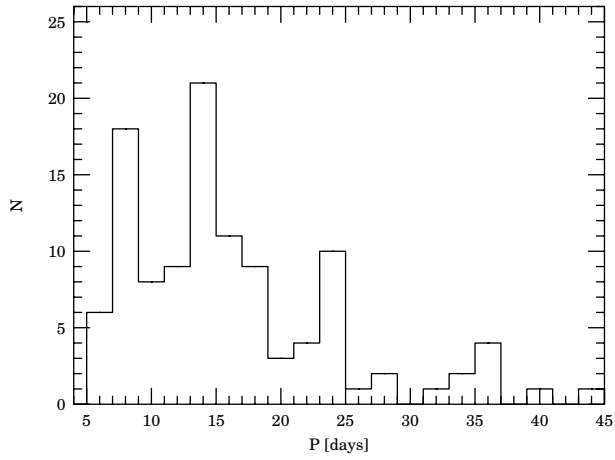}
\caption{Period distribution of Cepheids in NGC 300 from our catalog.} 
\end{figure}

\begin{figure}[b]
\vspace*{15 cm}
\includegraphics{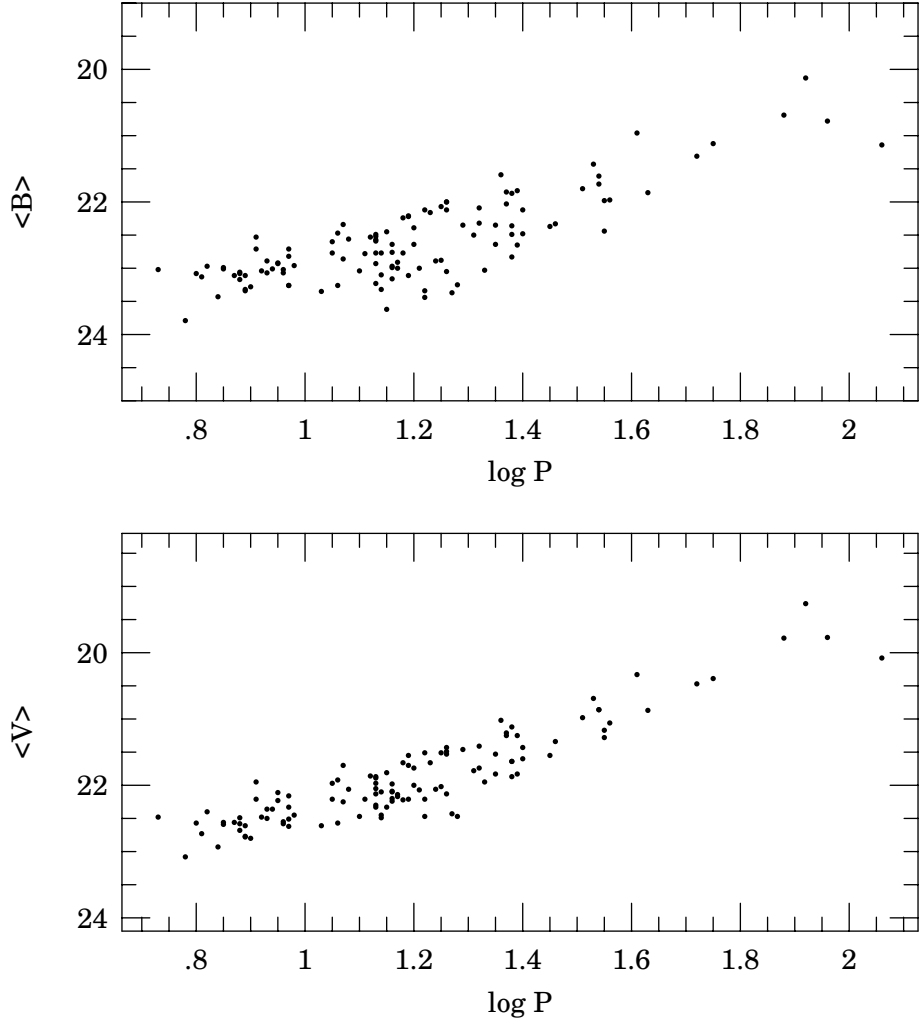}
\caption{PL relations for B and V magnitudes from the Cepheids in
NGC 300 detected in our survey. No absorption corrections have been applied to 
the mean magnitudes.}
\end{figure}

\end{document}